# THE MATHEMATICAL PHYSICAL EQUATIONS SATISFIED BY RETARDED AND ADVANCED GREEN'S FUNCTIONS

Huai-Yu Wang[a]

Department of Physics, Tsinghua University, Beijing, 100084 China

**Abstract:** In mathematical physics, time-dependent Green's functions (GFs) are the solutions of differential equations of the first and second time derivatives. Habitually, the time-dependent GFs are Fourier transformed into the frequency space. Then, analytical continuation of the frequency is extended to below or above the real axis. After inverse Fourier transformation, retarded and advanced GFs can be obtained, and there may be arbitrariness in such analytical continuation. In the present work, we establish the differential equations from which the retarded and advanced GFs are rigorously solved. The key point is that the derivative of the time step function is the Dirac $\delta$ function plus an infinitely small quantity, where the latter is not negligible because it embodies the meaning of time delay or time advance. The retarded and advanced GFs defined in this paper are the same as the one-body GFs defined with the help of the creation and destruction operators in many-body theory. There is no way to define the causal GF in mathematical physics, and the reason is given. This work puts the initial conditions into differential equations, thereby paving a way for solving the problem of why there are motions that are irreversible in time.

[a] wanghuaiyu@mail.tsinghua.edu.cn

**Résumé:** En physique mathématique, les fonctions de Green (GF) dépendant du temps sont les solutions des équations différentielles des dérivées première et seconde du temps. Habituellement, les GFs dépendant du temps sont transformées de Fourier dans l'espace des fréquences. Puis la fréquence est prise de suite analytique au-dessous ou au-dessus de l'axe réel. Après transformation de Fourier inverse, des GFs retardés et avancés peuvent être obtenus. Il peut y avoir de l'arbitraire dans une telle poursuite analytique. Dans ce travail, les équations différentielles pour les GFs retardés et avancés sont établies. Les GFs retardés et avancés sont résolus rigoureusement à partir des équations. Le point clé est que la dérivée de la fonction de pas de temps est la fonction de Dirac $\delta$ plus une quantité infiniment petite, où cette dernière n'est pas négligeable car elle incarne le sens de retard ou d'avance dans le temps. Les GFs retardés et avancés définis dans cet article sont les mêmes que les GFs à un corps définis à l'aide des opérateurs de création et de destruction dans la théorie à plusieurs corps. Les auteurs donnent la raison qu'il n'y a aucun moyen de définir le GF causal en physique mathématique. Ce travail met les conditions initiales dans les équations différentielles, ce qui ouvre la voie à la résolution du problème pourquoi il y a des mouvements qui sont irréversibles dans le temps.

## I. INTRODUCTION

In mathematical physics, there is no such differential equation that is satisfied by the retarded or advanced Green's function (GF) but merely differential equations for time-dependent GFs. After the GFs are solved, the skill of analytic continuation is utilized to produce the retarded or advanced GFs. In this work, we establish the equations for the retarded and advanced GFs and stress their physical meaning.

First of all, we present the differential equations of the first and second time derivatives that the time-dependent GFs satisfy in mathematical physics.

The first time derivative equation is[1–9]

$$(i\hbar \frac{\partial}{\partial t} - H(\mathbf{r}))G(\mathbf{r},\mathbf{r}';t,t') = i\hbar \delta(\mathbf{r}-\mathbf{r}')\delta(t-t'). \tag{1}$$

The second time derivative equation is[1,2,5,10–16]

$$(-\frac{\partial^2}{\partial t^2} - H(\mathbf{r}))G(\mathbf{r},\mathbf{r}';t,t') = \delta(\mathbf{r}-\mathbf{r}')\delta(t-t'). \tag{2}$$

In Eqs. (1) and (2), the $H$ operators are independent of time derivatives. Generally, $H$ contains the derivatives with respect to space coordinates of the second order.

In Eq. (1), $H$ is usually the Hamiltonian of the system under consideration. If the $H$ in (1) is the kinetic energy operator of the Schrödinger equation, the equation is thought to determine the propagation of a free particle[2,3] such that the solution is called the propagator.[4,8] Some[7] believe that both the retarded propagator and advanced propagator satisfy Eq. (1). If the $H$ in (1) is the Hamiltonian of the Dirac equation for a free particle, the solution is also called a propagator[2,5,6,8,9] and is named the Feynman propagator. The propagator is also expressed as a form of transition amplitude.[4,7]

There are two simplest and most important forms of $H$ in (2). One is the case in electrodynamics.[1,5,10,11,13,14,16]

$$(-\frac{1}{c^2}\frac{\partial^2}{\partial t^2} + \nabla_r^2)G(\mathbf{r},\mathbf{r}';t,t') = \delta(\mathbf{r}-\mathbf{r}')\delta(t-t'). \tag{3}$$

The $c$ in (3) is a constant, and in the case of vacuum electrodynamics, it is simply the light speed. The other is the case of the Klein-Gordon equation.[2,6,8,12,15,18,19]

$$(-\frac{\partial^2}{\partial t^2} + \nabla_r^2 - m^2)G(\mathbf{r},\mathbf{r}';t,t') = \delta(\mathbf{r}-\mathbf{r}')\delta(t-t'). \tag{4}$$

When discussing the case of (4), we set the unit $c = \hbar = 1$. In quantum electrodynamics (QED), the solution of Eq. (4) is also called the Feynman propagator, and is denoted by $\Delta_F$.[5,7,9,15] The solution in (3) is also called the photon propagator.[5,7]

In the theory of many-body GFs in condensed matter, there are definitions of the retarded, advanced and causal GFs.[1,20–22] They are defined by the ensemble average of the pairs of creation and annihilation operators of particles. In [22], some cases of the averages were listed, where the simplest was that in a vacuum state of the operator pairs of a non-interactive system. The GFs in the simplest case are termed one-body GFs.

Thus, there are actually two types of definitions of GFs, those in mathematical physics and those in many-body theory. The former are the solutions of Eqs. (1)-(4), which do not reflect the physical processes of the creation and annihilation of particles. The latter are closely related to the creation and annihilation of particles. In this work, we discuss the GFs in mathematical physics, and they may be compared with one-body GFs in many-body theory where possible.

In quantum mechanics (QM) and QED, as long as the creation and annihilation of particles are not involved, we always solve differential equations, so that merely the GFs in mathematics are concerned.

In QM, the default is usually that the solutions solved from a differential equation are time delaying unless otherwise specified. In classical electrodynamics, one also often discusses time-delaying solutions. In QED, the concept of time retard and advance are more frequently involved even without touching the creation and annihilation of particles. Therefore, we have to consider the time delay and advance in mathematical physics.

However, the solutions obtained from Eqs. (1)-(4) are not of the meanings of time delay and advance. In order to raise the meanings, the following procedure is carried out. The GF's time Fourier transformation is taken.

$$G(\bm{r},\bm{r}';t,t') = \frac{1}{2\pi}\int d\omega e^{-i\omega(t-t')} g(\bm{r},\bm{r}';\omega).\qquad(5)$$

Equation (5) is substituted into any of Eqs. (1)-(4) to obtain the expressions of $g(\bm{r},\bm{r}';\omega)$. The problem is that, on the real $\omega$ axis, the $g(\bm{r},\bm{r}';\omega)$ may have poles and hence cannot be defined. In this case, analytic continuation of $g(\bm{r},\bm{r}';\omega)$ occurs in the $\omega$ plane.[1] Usually, when deviating from the real $\omega$ axis, an infinitely small imaginary part $i\eta$ is attached, where $\eta = 0^+$ denotes a positive infinitesimal. After the inverse time Fourier transformation is made, the retarded or advanced GF is gained. The retarded and advanced GFs are denoted by $G^R$ and $G^A$, respectively. In this paper, the superscripts R and A respectively denote time delay and advance.

One thing is noted: in executing the analytic continuation, there may be some arbitrariness. A typical example is that when the square of the frequency appears in the denominator, there are two possible ways to conduct the continuation.

One is that the $i\eta$ is added to the frequency.[10,b]

---

[b]In [1,11,13,14,16], the manipulation was the same as Eq. (6) although not explicitly written.

$$\frac{1}{\omega^2 - c^2 k^2} \to \frac{1}{(\omega + i\eta)^2 - c^2 k^2}. \tag{6}$$

The other is that the $i\eta$ is added to the frequency square.[5,7,8,15]

$$\frac{1}{\omega^2 - (E_k/\hbar)^2} \to \frac{1}{\omega^2 - (E_k/\hbar)^2 + i\eta}. \tag{7}$$

It is believed that how to do the continuation relies on the physical intuitive. For instance, Eq. (6) "is meaningless without some rule as to how to handle the singularities. The rule cannot come from the mathematics. It must come from physical considerations."[10] Nevertheless, every person may have his own point of view in conducting the continuation. For example, when the dispersion relation is $E_k^2 = k^2 + m^2$, people may add the imaginary part to the mass in the denominator of Eq. (7) in the way of $m \to m - i\eta$.[2,5,6,8]

This shows that discrepancy might be raised when different people conduct the continuation based on their own viewpoints. In the present authors' opinion, a rigorous mathematical proof ought to be provided for an analytical continuation. Under uniform mathematical manipulation, the correct one in (6) and (7) will be chosen.

Since the GFs $G$ solved from Eqs. (1)-(4) do not possess the meaning of time delay and advance, they can be merely regarded as auxiliary functions.[c] They are still simply called the GF here to maintain the habit, and it is the $G^R$ and $G^A$ that have real physical meaning.

In this paper, the differential equations that $G^R$ and $G^A$ satisfy are proposed, and the explicit relations between each of them and the $G$ are given. When this is done, the arbitrariness reflected by Eqs. (6) and (7) will be eliminated.

The content is arranged as follows. Section II discusses the first order equations. We start from the fundamental equations of QM. In QM, people usually deal with time-retarded solutions. We specify the equation that the retarded wave functions should follow. Then, the equation for the corresponding retarded GF is suggested. In Section III the second order equations are discussed. Section IV is some discussion. It is stressed that the equations for $G$ are of invariance of time translation and time inversion, but those for the retarded and advanced GFs are not. Revealing this discrepancy paves the way to determining why there are irreversible processes. Finally, our conclusions are detailed in Section V. Appendix shows the result of the derivative of the step function, where an infinitesimal should be retained.

---

[c]The auxiliary Green's function $G$ was denoted by different symbols in the literature. For examples, it was $\tilde{g}$ in [1], $D$ in [13,24], and $\Delta$ in [15,19]. Please note that the poles of this function are on the real axis. Therefore, it is not defined on the axis. In the literature, different names were endowed to the function $G$ in the ranges $t > t'$ and $t < t'$. The $G(t > t')$ and $G(t < t')$, for instances, were respectively denoted by $g^>$ and $g^<$ in [1], $\Delta^+$ 和 $\Delta^-$ in [5], $G^{(+)}$ and $G^{(-)}$ in [7], $\Delta^{(+)}$ and $\Delta^{(-)}$ in [15], $\Delta_+$ and $\Delta_-$ in [19].

## II. THE FIRST ORDER EQUATIONS

The fundamental equations of QM, Schrödinger equation and Dirac equation, have of the following form:

$$(i\hbar \frac{\partial}{\partial t} - H)\psi(\boldsymbol{r},t) = 0. \tag{8}$$

Its formal solution can be written as

$$\psi(\boldsymbol{r},t) = \exp(-\frac{i}{\hbar} \int_{t'}^{t} dt_1 H)\psi(\boldsymbol{r},t'). \tag{9}$$

This expression shows that as soon as the wave function at the initial instant is known, it can be evaluated at any later time.

Here, we have actually mentioned two concepts: initial condition and time delay. The initial condition is necessary to solve a differential equation with the first order time derivative. In Eq. (9), the solution contains the initial value $\psi(\boldsymbol{r},t')$. The phrase "at any later time" means a retarded solution.

In fact, Eqs. (8) and (9) themselves do not reflect time delay because, in these equations, the time $t$ can be either after or before the time $t'$. To be rigorous, the retarded solution should be explicitly indicated mathematically. Some authors,[4,5,7,8,23] usually in discussing scattering problems, have noticed this: a factor of time step function $\theta(t-t')$ ought to be attached, although they did not address this through derivation of formulas. This factor demonstrates that after we set the instant $t'$ as the starting point of time, we merely consider what will happen at time $t \geq t'$.

Hence, the retarded solution should be expressed by

$$\psi^R(\boldsymbol{r},t) = \theta(t-t')\exp(-\frac{i}{\hbar}\int_{t'}^{t} dt_1 H)\psi^R(\boldsymbol{r},t'). \tag{10}$$

We think that as long as a solution is for time delay, the factor $\theta(t-t')$ must be included. After taking the time derivative with respect to Eq. (10), we obtain

$$(i\hbar\frac{\partial}{\partial t} - H)\psi^R(\boldsymbol{r},t) = i\hbar\frac{\partial \theta(t-t')}{\partial t}\psi^R(\boldsymbol{r},t'). \tag{11}$$

Please note that we do not replace $\frac{\partial \theta(t-t')}{\partial t}$ with $\delta(t-t')$ because there is an infinitesimal quantity that should not be abandoned (see Appendix). It is this infinitesimal that embodies the time delay. As a comparison, the $\delta(t-t')$ functions on the right-hand sides of Eqs. (1)-(4) reflect neither retard nor advance. On the right-hand side of (11), the factor $\frac{\partial \theta(t-t')}{\partial t}$ reflects time retard, and the $\psi^R(\boldsymbol{r},t')$ gives the initial condition.[d] The form of (11) unambiguously describes the evolution of the wave function toward the future.

Equation (8), as a homogeneous differential equation with the first order time derivative, enables us to directly input the initial condition into the differential equation to obtain (11). When $t > t'$, Eqs. (11) and (8) are identical and differ only at the instant $t = t'$.

Now, let us solve (11). Because the solution $\psi^R(\boldsymbol{r},t)$ embodies the time delay, it must contain the factor $\theta(t-t')$. Therefore, we let

$$\psi^R(\boldsymbol{r},t') = \theta(t-t')\psi(\boldsymbol{r},t'). \tag{12}$$

Equation (12) is substituted into (11) to get

$$\theta(t-t')(i\hbar\frac{\partial}{\partial t} - H(\boldsymbol{r}))\psi(\boldsymbol{r},t) + i\hbar\psi(\boldsymbol{r},t)\frac{\partial}{\partial t}\theta(t-t')$$
$$= i\hbar\frac{\partial \theta(t-t')}{\partial t}\psi^R(\boldsymbol{r},t'). \tag{13}$$

We let the first term on the left-hand side equal zero. Then, we have the following differential equation and initial condition.

$$(i\hbar\frac{\partial}{\partial t} - H(\boldsymbol{r}))\psi(\boldsymbol{r},t) = 0, \quad t > t' \tag{14a}$$

$$\psi(\boldsymbol{r},t)|_{t=t'} = \psi^R(\boldsymbol{r},t'). \tag{14b}$$

Thus, Eq. (11) automatically contains equation (14a) and the initial condition (14b). Please note that the time range of the equation is indicated in (14a). In (14b), the initial condition is that the time $t$ approaches $t'$ on the $t > t'$ side.

In the same way, we can derive the equation describing the evolution of the wave function toward the past, which is not explicitly written here.

How to solve Eq. (14) is introduced in QM textbooks. Here we solve it following

---

[d]The term on the right-hand side of (11) may have further physical meanings, which will be investigated in our future work, see the last three paragraphs in Section IV.

a standard routine under the condition that the Hamiltonian $H$ is independent of time.

We assume that with given boundary conditions, the eigenvalues $\{E_n\}$ and eigenfunction set $\{\varphi_n(\boldsymbol{r})\}$ are solved.

$$H(\boldsymbol{r})\varphi_n(\boldsymbol{r}) = E_n \varphi_n(\boldsymbol{r}). \tag{15}$$

The eigenfunction set is complete:

$$\delta(\boldsymbol{r}-\boldsymbol{r}') = \sum_n \varphi_n(\boldsymbol{r})\varphi_n^*(\boldsymbol{r}'). \tag{16}$$

The solution $\psi(\boldsymbol{r},t)$ is expanded by the complete set. The solution of Eq. (14) is readily derived as

$$\psi(\boldsymbol{r},t) = \sum_n c_n \varphi_n(\boldsymbol{r}) e^{-iE_n(t-t')/\hbar}, \tag{17}$$

where the coefficients $\{c_n\}$ are to be determined by the initial condition (14b):

$$\sum_n c_n \varphi_n(\boldsymbol{r}) = \psi^R(\boldsymbol{r},t'). \tag{18}$$

The first example is a free particle of the Schrödinger equation. Its Hamiltonian, eigenvalues and eigenfunctions are respectively

$$H = -\frac{\hbar^2}{2m}\nabla^2, \quad E_k = \frac{\hbar^2 k^2}{2m}, \tag{19}$$

and

$$\varphi_k(\boldsymbol{r}) = \frac{1}{(2\pi)^{3/2}} e^{i\boldsymbol{k}\cdot\boldsymbol{r}}. \tag{20}$$

If at the initial instant, the wave function is a plane wave, $\psi^R(\boldsymbol{r},t') = \varphi_k(\boldsymbol{r}) = \frac{1}{(2\pi)^{3/2}} e^{i\boldsymbol{k}\cdot\boldsymbol{r}}$, then we get $c_k = 1$ from (18). At any later time, the wave function is still the plane wave: $\psi(\boldsymbol{r},t) = \frac{1}{(2\pi)^{3/2}} e^{i\boldsymbol{k}\cdot\boldsymbol{r}} e^{-iE_k(t-t')/\hbar}$. If at the initial instant, the wave function is the $\delta(\boldsymbol{r})$ function located at the origin, $\psi^R(\boldsymbol{r},0) = \delta(\boldsymbol{r})$, then the coefficients are $c_k = \frac{1}{(2\pi)^{3/2}}$, and the wave function at any later time is

$\psi(\boldsymbol{r},t) = \frac{1}{(4\pi i(t-t'))^{3/2}} e^{imr^2/2\hbar(t-t')}$. This solution collapses with time.

If a Schrödinger particle is in a one-dimensional square potential well within the interval [0,$a$], the eigenvalues and eigenfunctions are respectively $E_n = \frac{\pi^2 \hbar^2 n^2}{2ma^2}$ and $\varphi_n(x) = \sqrt{\frac{2}{a}} \sin \frac{n\pi x}{a}$. Suppose that the initial wave function is $\psi^R(x,t') = \varphi_l(x) = \sqrt{\frac{2}{a}} \sin \frac{l\pi x}{a}$; then, $c_l = 1$ and the solution is $\psi(x,t) = \varphi_l(x) e^{-iE_l(t-t')/\hbar} = \sqrt{\frac{2}{a}} \sin \frac{l\pi x}{a} e^{-iE_l(t-t')/\hbar}$. Suppose that the initial wave function is the $\delta$ function at the well center, $\psi^R(x,0) = \delta(x-a/2)$; then the coefficients are $c_{2m} = 0, c_{2m+1} = i\sqrt{2/a}$.

The third example is the one-dimensional harmonic oscillator. Its Hamiltonian is

$$H = -\frac{\hbar^2}{2m}\nabla^2 + \frac{1}{2}m\omega^2 x^2 \tag{21}$$

and the eigenvalues are

$$E_n = (n+\frac{1}{2})\hbar\omega, n = 0,1,2,\cdots. \tag{22}$$

The eigenfunctions are

$$\varphi_n(x) = N_n e^{-\alpha^2 x^2/2} H_n(\alpha x), \tag{23}$$

where $\alpha = \sqrt{\frac{m\omega}{\hbar}}, N_n = (\frac{\alpha}{\sqrt{\pi}2^n n!})^{1/2}$ and $H_n(x)$ is the Hermite polynomial of $n$ order. If at the initial instant $\psi^R(x,t') = \varphi_n(x)$, then $c_n = 1$ and all other values of $c_n$ are zero. At any later time, the wave function is $\psi^R(x,t) = \varphi_n(x)e^{-iE_n(t-t')/\hbar}$. If the initial wave function is the $\delta(x)$ function located at the origin, $\psi^R(x,0) = \delta(x)$, then

$$c_{2n} = (-1)^n \frac{1}{2^n n!}(\frac{\alpha(2n)!}{\sqrt{\pi}})^{1/2}, c_{2n+1} = 0.$$

We turn to investigate the retarded GF corresponding to Eq. (11). We define the retarded GF that satisfies the following differential equation with the first order time derivative.

$$(i\hbar\frac{\partial}{\partial t} - H(\mathbf{r}))G^R(\mathbf{r},\mathbf{r}';t,t') = i\hbar\delta(\mathbf{r}-\mathbf{r}')\frac{\partial}{\partial t}\theta(t-t'). \tag{24}$$

Equation (24) can be compared with (1). The only difference between them is that the factor $\delta(t-t')$ in (1) is replaced by $\frac{\partial \theta(t-t')}{\partial t}$ in (24). Similarly to (11), the factor $\frac{\partial \theta(t-t')}{\partial t}$ explicitly demonstrates that Eq. (24) produces retarded solutions.

The procedure of solving Eq. (24) is the same as that solving (11). There must be a factor of time step function $\theta(t-t')$ in $G^R(r,r';t,t')$. We let

$$G^R(r,r';t,t') = \theta(t-t')G(r,r';t,t'). \tag{25}$$

When it is substituted into (24), we achieve

$$\theta(t-t')(i\hbar\frac{\partial}{\partial t} - H(r))G(r,r';t,t') + i\hbar G(r,r';t,t')\frac{\partial}{\partial t}\theta(t-t')$$
$$= i\hbar\delta(r-r')\frac{\partial}{\partial t}\theta(t-t'). \tag{26}$$

We let the first term on the left-hand side be zero. It follows that

$$(i\hbar\frac{\partial}{\partial t} - H(r))G(r,r';t,t') = 0 \tag{27a}$$

and

$$G(r,r';t,t')|_{t=t'} = \delta(r-r'). \tag{27b}$$

The solution $G$ solved from (27a) under the initial condition (27b) is actually an auxiliary function.[c] As long as the auxiliary function is known, the retarded GF $G^R(r,r';t,t')$ can be written down in terms of Eq. (25). Please note that in (27b), the time $t$ approaches $t'$ on the $t > t'$ side. This initial condition has been previously mentioned,[3] and it indicates that the function $G$ is of the form of the weak convergence of the $\delta$ function at the limit $t - t' \to 0^+$.[24]

The solution of Eq. (11) can be expressed using the retarded GF solved from (24). This is achieved by acting the inverse operator $(i\hbar\frac{\partial}{\partial t} - H)^{-1}$ on both sides of (11) and (24).

$$\psi^R(r,t) = (i\hbar\frac{\partial}{\partial t} - H)^{-1}i\hbar\frac{\partial \theta(t-t')}{\partial t}\psi^R(r,t')$$
$$= \int dr'(i\hbar\frac{\partial}{\partial t} - H)^{-1}i\hbar\frac{\partial \theta(t-t')}{\partial t}\delta(r-r')\psi^R(r',t') \tag{28}$$
$$= \int dr' G^R(r,r';t,t')\psi^R(r',t').$$

This formula demonstrates that the $G^R(\boldsymbol{r},\boldsymbol{r}';t,t')$ plays such a role that it helps to propagate one state to another through all possible intermediate paths.

Since $G^R(\boldsymbol{r},\boldsymbol{r}';t,t')$ has a factor $\theta(t-t')$, see (25), the solution (28) automatically has the form of (12). Equations (12), (25), and (28) are self-consistent.

Equation (28) can be repeatedly iterated. A single round of iteration yields

$$\psi^R(\boldsymbol{r},t) = \int d\boldsymbol{r}' \int d\boldsymbol{r}_1 G^R(\boldsymbol{r},\boldsymbol{r}';t,t') G^R(\boldsymbol{r}',\boldsymbol{r}_1;t',t_1) \psi^R(\boldsymbol{r}_1,t_1). \tag{29}$$

This formula has appeared before, see [2,3], where the retarded GF was denoted by letter $K$ and its expression was written as

$$G(\boldsymbol{r},\boldsymbol{r}';t,t') = -i\sum_n \varphi_n(\boldsymbol{r})\varphi_n^*(\boldsymbol{r}') e^{-iE_n(t-t')/\hbar}. \tag{30}$$

The expression is solved from Eq. (27) through the following procedure: under the conditions that the Hamiltonian is independent of time and Eqs. (15) and (16) stand, the function $G$ was expanded by the complete set solved from (15), and then substituted into (27a); finally, the condition (27b) and the identity (16) are employed. Equation (30) has been given in the literature[1-5] and is called the retarded GF or propagator.

We stress again that the $G$ expressed by (30) is simply an auxiliary function. The real retarded GF, with the help of (25), is

$$G^R(\boldsymbol{r},\boldsymbol{r}';t,t') = -i\theta(t-t')\sum_n \varphi_n(\boldsymbol{r})\varphi_n^*(\boldsymbol{r}') e^{-iE_n(t-t')/\hbar}. \tag{31}$$

The $G^R$, as the solution of (24), is physically meaningful. People may multiply a factor $\theta(t-t')$ to (30) directly,[5,15] or default this factor without showing it explicitly.[2,3]

Now, we enumerate some of the simplest examples.

In the case of a Schrödinger free particle, Eqs. (19) and (20) are input into (30) to obtain the auxiliary function,[5] $G(\boldsymbol{r},\boldsymbol{r}';t,t') = \dfrac{m^{3/2} e^{i(\boldsymbol{r}-\boldsymbol{r}')^2 m/2\hbar(t-t')}}{(2\pi i\hbar(t-t'))^{3/2}}$. This is well known, and the exponent contains the classical action $S_C = \dfrac{m(\boldsymbol{r}-\boldsymbol{r}')^2}{2(t-t')}$.

In the case of a one-dimensional harmonic oscillator, Eqs. (21)-(23) are input into (30) to obtain

$$G(x,x';t,t') = \frac{\sqrt{m\omega}}{\sqrt{2\pi\hbar i \sin\omega(t-t')}} \exp[im\omega \frac{(x^2+x'^2)\cos\omega(t-t') - 2xx'}{2\hbar\sin\omega(t-t')}].$$

This was thought of as the propagator of the oscillator and has been derived in various ways.[25–30]

We take the Fourier transformation of the $G^R(\boldsymbol{r},\boldsymbol{r}';t,t')$.

$$G^R(\mathbf{r},\mathbf{r}';t,t') = \int d\mathbf{k} e^{i\mathbf{k}\cdot(\mathbf{r}-\mathbf{r}')} e^{-i\omega(t-t')} G^+(\mathbf{k},\omega). \tag{32}$$

The convolution of the time Fourier components of the two factors in (25) results in

$$G^+(\mathbf{r},\mathbf{r}';\omega) = \frac{1}{2\pi}\int_{-\infty}^{\infty} d\omega' \frac{g(\mathbf{r},\mathbf{r}';\omega')}{\omega-\omega'+i\eta}, \tag{33}$$

where the $g(\mathbf{r},\mathbf{r}';\omega)$ is calculated by Eq. (5). We have denoted

$$G^{\pm}(\mathbf{r},\mathbf{r}';\omega) = G(\mathbf{r},\mathbf{r}';\omega\pm i\eta). \tag{34}$$

Their spatial Fourier transformations are

$$G^{\pm}(\mathbf{k},\omega) = G(\mathbf{k},\omega\pm i\eta). \tag{35}$$

In the present case, the infinitesimal imaginary part in the denominator of (33) is automatically determined by Fourier transformation of the factor $\theta(t-t')$ in (25), see Eq. (A5), and not by artificial analytical continuation. This means that mathematical rigorousness is guaranteed. The positive imaginary part $i\eta$ represents the time delay because it comes from the factor $\theta(t-t')$.

From Eq. (30), we have

$$g(\mathbf{r},\mathbf{r}';\omega) = \sum_n \varphi_n(\mathbf{r})\varphi_n^*(\mathbf{r}')\delta(\omega-E_n/\hbar). \tag{36}$$

It follows that

$$G^+(\mathbf{r},\mathbf{r}';\omega) = \sum_n \frac{\varphi_n(\mathbf{r})\varphi_n^*(\mathbf{r}')}{\omega-E_n/\hbar+i\eta}. \tag{37}$$

This is the same as in one-body GF in many-body theory.[22]

In the case of a Schrödinger free particle, Eqs. (19) and (20) are substituted into (36) and (37) to get

$$g(\mathbf{k},\omega) = \delta(\omega-E_k/\hbar) \tag{38}$$

and

$$G^+(\mathbf{k},\omega) = \frac{1}{\omega-E_k/\hbar+i\eta}. \tag{39}$$

It is noted that Eq. (36) is simply the Fourier component of the auxiliary function (30), but not that of the retarded GF (31). Therefore, no one utilizes Eq. (30) as the Fourier transformation of the retarded GF in the literature.

For a relativistic particle with spin 0, the energy spectrum has positive and negative branches.

$$E_{(\pm)\mathbf{k}} = \pm\sqrt{m^2c^4+c^2\hbar^2\mathbf{k}^2} = \pm E_\mathbf{k}. \tag{40}$$

The wave functions are still (20). It is emphasized that both the branches should be

contained.[e] The retarded GF is

$$G^{R}(\boldsymbol{r},\boldsymbol{r}';t,t') = -\mathrm{i}\theta(t-t')\frac{1}{(2\pi)^3}\int d\boldsymbol{k} e^{i\boldsymbol{k}\cdot(\boldsymbol{r}-\boldsymbol{r}')}(e^{-iE_k(t-t')/\hbar} + e^{iE_k(t-t')/\hbar}). \quad (41)$$

We denote

$$\Delta_{(+)+} = \int \frac{d\boldsymbol{k}}{(2\pi)^3}\frac{e^{i\boldsymbol{k}\cdot(\boldsymbol{r}-\boldsymbol{r}')}}{\omega - E_k/\hbar + i0^+}, \quad \Delta_{(-)+} = \int \frac{d\boldsymbol{k}}{(2\pi)^3}\frac{e^{i\boldsymbol{k}\cdot(\boldsymbol{r}-\boldsymbol{r}')}}{\omega + E_k/\hbar + i0^+}. \quad (42)$$

Then the Fourier component is

$$G^{+}(\boldsymbol{r},\boldsymbol{r}';\omega) = \Delta_{(+)+} + \Delta_{(-)+} = \frac{2\omega}{(2\pi)^3}\int d\boldsymbol{k}\frac{e^{i\boldsymbol{k}\cdot(\boldsymbol{r}-\boldsymbol{r}')}}{\omega^2 - (E_k/\hbar)^2 + \omega i0^+}. \quad (43)$$

This form of the infinitesimal imaginary part in the denominator has been earlier noticed.[6] Equation (43) is consistent with (6) instead of (7). We emphasize again that (43) is rigorously derived, and we do not implement any artificial manipulation. The retarded GF necessarily contains a factor $\theta(t-t')$ from which the infinitesimal imaginary part of the frequency is derived. The poles are certainly below the real $\omega$ axis for both the positive and negative energy branches.

The above discussion reveals that even if the analytical continuation of the Fourier component is conducted based on some physical consideration, a rigorous derivation is still needed, just as in the present work.

The advanced GF $G^{A}(\boldsymbol{r},\boldsymbol{r}';t,t')$ can be investigated in the same way. The $G^{A}(\boldsymbol{r},\boldsymbol{r}';t,t')$ is defined as the solution satisfying the following equation:

$$(i\hbar\frac{\partial}{\partial t} - H)G^{A}(\boldsymbol{r},\boldsymbol{r}';t,t') = -i\hbar\frac{\partial\theta(t'-t)}{\partial t}\delta(\boldsymbol{r}-\boldsymbol{r}'). \quad (44)$$

It is related to the GF by

$$G^{A}(\boldsymbol{r},\boldsymbol{r}';t,t') = -\theta(t'-t)G(\boldsymbol{r},\boldsymbol{r}';t,t'). \quad (45)$$

Substitution of (45) into (44) leads to

$$(i\hbar\frac{\partial}{\partial t} - H(\boldsymbol{r}))G(\boldsymbol{r},\boldsymbol{r}';t,t') = 0 \quad (46a)$$

and

$$G(\boldsymbol{r},\boldsymbol{r}';t,t')|_{t=t'} = \delta(\boldsymbol{r}-\boldsymbol{r}'). \quad (46b)$$

One can compare Eqs. (27) and (46), and see that their solutions are the same. Please note that in (46b), the time $t$ approaches $t'$ on the $t < t'$ side.

---

[e] Here we merely consider the cases that the particles do not have electric charge. The charged particles will be researched later.

If the Hamiltonian is independent of time and Eqs. (15) and (16) stand, the solved $G$ is again (30). The $G^A(r,r';t,t')$ is

$$G^A(r,r';t,t') = i\theta(t'-t)\sum_n \varphi_n(r)\varphi_n^*(r')e^{-iE_n(t-t')/\hbar}. \tag{47}$$

The time Fourier component is

$$G^-(r,r';\omega) = \frac{1}{2\pi}\int_{-\infty}^{\infty} d\omega' \frac{g(r,r';\omega')}{\omega - \omega' - i\eta} = \sum_n \frac{\varphi_n(r)\varphi_n^*(r')}{\omega - E_n/\hbar - i\eta}. \tag{48}$$

This is the same as that in one-body GF in many-body theory. The pole is above the real $\omega$ axis, the reflection of time advance. For a Schrödinger free particle, the spatial Fourier transformation is

$$G^-(k,\omega) = \frac{1}{\omega - E_k/\hbar - i\eta}. \tag{49}$$

For a relativistic particle with spin 0, $G^A(r,r';t,t')$ is

$$G^A(r,r';t,t') = i\theta(t'-t)\int \frac{dk}{(2\pi)^3} e^{ik\cdot(r-r')}(e^{-iE_k(t-t')/\hbar} + e^{iE_k(t-t')/\hbar}). \tag{50}$$

We denote

$$\Delta_{(-)+} = \int \frac{dk}{(2\pi)^3} \frac{e^{ik\cdot(r-r')}}{\omega - E_k/\hbar - i\eta}, \quad \Delta_{(-)-} = \int \frac{dk}{(2\pi)^3} \frac{e^{ik\cdot(r-r')}}{\omega + E_k/\hbar - i\eta}. \tag{51}$$

Then the Fourier component is the sum of the two terms.[6]

$$G^-(r,r';\omega) = \Delta_{(-)+} + \Delta_{(-)-} = \frac{2\omega}{(2\pi)^3} \int dk \frac{e^{ik\cdot(r-r')}}{\omega^2 - (E_k/\hbar)^2 - \omega i\eta}. \tag{52}$$

Equations (24) and (44) can also be applied to the case of the Dirac equation,[2,5] the solution of which was thought to determine the propagation of a Dirac particle.[2] All the discussions above are applicable to the case of the Dirac equation. One merely keeps in mind that the Hamiltonian, as well as the GFs, is of a matrix form, and the wave functions are of spinor forms. We do not discuss the case in detail here.

Now, we sum up the first term in (42) and the second term in (51) to get

$$\Delta_F = \Delta_{(+)+} + \Delta_{(-)-} = \frac{2\omega}{(2\pi)^3} \int dk \frac{e^{ik\cdot(r-r')}}{\omega^2 - (E_k/\hbar)^2 + i\eta}. \tag{53}$$

This is the same as the analytical continuation in Eq. (7), which shows that (7) is not solved from one equation but simply by piecing together. There is no such result in mathematical physics.

In the many-body theory, the causal GF is defined,[1,20,22] where the factors $\theta(t-t')$ and $\theta(t'-t)$ unmistakably indicate the time orders in the two terms. It is our opinion that in mathematical physics, there is no way to define the differential equation that the causal GF satisfies such that there is no concept of a causal GF.

It has been mentioned above that the solution of Eq. (1) is also called the Feynman propagator and denoted as $\Delta_F$. Let us review this concept. There are actually two definitions of the Feynman propagator in the textbooks of QE and quantum field theory. One is the solution of (1), a definition in mathematical physics. The other is using the creation and annihilation operators.[5,6,18] Some writers present both,[5,6,9] and think that the results from the two definitions are the same.

The propagator defined by the creation and annihilation operators coincides with the causal GF in many-body theory.[1,20–22] The two terms each contain the factors $\theta(t-t')$ and $\theta(t'-t)$ such that the poles of the two terms in the frequency space are below and above the real $\omega$ axis, just as in (53).

In mathematical physics, we have defined differential equations (24) and (44). Their solutions $G^R$ and $G^A$ explicitly reflect time delay and advance, respectively. The poles of the $G^R(\omega)$ are below the real axis, and those of the $G^A(\omega)$ above the axis. We are unable to establish a differential equation, for which the solution has two terms each having a pole, with one pole below and the other above the real axis, as (53).

In conclusion, the causal GF cannot be defined in mathematical physics because there is no equation it satisfies, despite the possibility to define it by means of the ensemble average of the creation and annihilation operators in many-body theory.

## III. THE SECOND ORDER EQUATIONS

There is a discrepancy in the properties of differential equations with the second derivatives compared with those of the first derivative. A remarkable feature of the former is that the solutions constitute a complete set under appropriate boundary conditions, whereas this feature is lacking in the equations with the first derivative. The physical meanings reflected by the first and second order equations can be different. For example, with the inclusion of the second order of spatial derivatives, a first order time derivative equation, as a diffusion-type equation, mainly describes the evolution or propagation of a state, while a second order time derivative may embody the distribution of a field generated from a source, and the field varies with time.

We define the retarded GF as satisfying the following differential equation with the second order time derivative.

$$(-\frac{1}{c^2}\frac{\partial^2}{\partial t^2} - H)G^R(\mathbf{r},\mathbf{r}';t,t') = -\frac{\partial \theta(t-t')}{\partial t}\delta(\mathbf{r}-\mathbf{r}'). \tag{54}$$

The operator $H$ does not contain a time derivative. The factor $\dfrac{\partial \theta(t-t')}{\partial t}$ on the right-hand side indicates the time retard. We let

$$G^R(\boldsymbol{r},\boldsymbol{r}';t,t') = \theta(t-t')G. \tag{55}$$

The function $G$ is required to meet the following initial conditions:

$$G|_{t=t'} = 0 \tag{56a}$$

and

$$\frac{\partial G}{\partial t}\bigg|_{t=t'} = c^2\delta(\boldsymbol{r}-\boldsymbol{r}'). \tag{56b}$$

In (56b), the time $t$ approaches $t'$ on the $t>t'$ side. The initial condition (56a) means that before the initial instant, there is no field source, and (56b) means that at the instant $t'$, there appears a point source at $\boldsymbol{r}'$. After that, this point source remains in its position with constant strength.

From (55) and (56), it is known that

$$\frac{\partial \theta(t-t')}{\partial t}G = 0. \tag{57}$$

Substituting (55) into (54) and using (56) and (57), we obtain

$$(-\frac{1}{c^2}\frac{\partial^2}{\partial t^2} - H)G = 0. \tag{58}$$

Equations (56)-(58) are for solving $G$, which is an auxiliary function.[c] The relationship (55) was earlier noted,[13,15] and the initial conditions (56) also mentioned.[1,9,15,19,31]

The physical meaning of Eq. (54) is as follows: at the instant $t'$ at the position $\boldsymbol{r}'$, there appears a point source; at any later time the strength and position of the source remain unchanged; the source stimulates a field in space; the equation determines the field distribution in space at any time $t>t'$. The two factors on the right hand side of (54) demonstrate the inequivalence of time and space--in solving the equation, the $\boldsymbol{r}$ can be any point in the specified room, often the whole space, while $t$ cannot be any time and is restricted to the future. The physically real solutions are those that are retarded. The solutions satisfying (54) instead of (2) represent the real movement.

We solve Eq. (58) under the initial conditions (56). Suppose that the operator $H$ is independent of time and that Eqs. (15) and (16) stand. The procedure for solving is similar to that of Eq. (30): the function $G$ was expanded by the complete set solved from (15), and then substituted into (58); finally, the conditions (56) and the identity (16) are employed. The result is that

$$G = c\sum_n \varphi_n(\boldsymbol{r})\varphi_n^*(\boldsymbol{r}')\frac{\sin\sqrt{E_n}c(t-t')}{\sqrt{E_n}}. \tag{59}$$

The operator $H$ in (54) is not a Hamiltonian, and its eigenvalues $E_n$ are not the energies of the researched system. In [1], by assigning the integral paths above or below the real $\omega$ axis, Eq. (59) was also obtained and denoted $\tilde{g}$. Here it is revealed that the $G$ is actually the auxiliary GF satisfying (58) and (56) without need of artificially assigning integral paths.

In the case of the scalar and vector potentials of electromagnetic field, i.e., Eq. (3),

$$H = -\nabla_r^2, \quad \varphi_k(r) = e^{-ik\cdot r} \quad \text{and} \quad E = k^2. \tag{60}$$

Equation (59) is simplified as

$$\begin{aligned}G &= c\sum_k e^{-ik\cdot(r-r')} \frac{\sin kc(t-t')}{k} \\ &= \frac{1}{4\pi|r-r'|}[\delta(t-t'+\frac{|r-r'|}{c}) - \delta(t-t'-\frac{|r-r'|}{c})].\end{aligned} \tag{61}$$

The retarded GF is

$$G^R = \theta(t-t')\frac{1}{4\pi|r-r'|}\delta(t-t'-\frac{|r-r'|}{c}). \tag{62}$$

This is the retarded potential in a vacuum generated by the point source located at the origin.

In the literature, when (61) is obtained,[11,16] it is reformed, by dropping the second term, to become the retarded GF. The dropping manipulation is actually multiplying a factor $\theta(t-t')$ to the $G$, just as in (55). Here, we rigorously obtain (62) from the equation (54) without any artificial manipulation.

Another special case of (54) is the Klein--Gordon equation, i.e., Eq. (4).

$$H = -\nabla_r^2 + m^2, \quad \varphi_k(r) = e^{-ik\cdot r}, \quad E_n \to E_k^2, E_k = \pm\sqrt{k^2+m^2}. \tag{63}$$

Equation (63) is substituted in (59) to get

$$G(r-r';t-t') = \frac{1}{(2\pi)^3}\int dk e^{ik\cdot(r-r')}\frac{1}{2iE_k}(e^{iE_k(t-t')} - e^{-iE_k(t-t')}). \tag{64}$$

This expression can be reformed into the following way.

$$\begin{aligned}&G(r-r';t-t') \\ &= \frac{1}{(2\pi)^3 i}\int dk e^{ik\cdot(r-r')}\frac{1}{2E_k}\frac{1}{2\pi}\int d\omega(\frac{e^{-i\omega(t-t')}}{\omega+E_k+i\eta} - \frac{e^{-i\omega(t-t')}}{\omega-E_k+i\eta}) \\ &= -\frac{1}{(2\pi)^4 i}\int d\omega\int dk \frac{e^{ik\cdot(r-r')}e^{-i\omega(t-t')}}{\omega^2 - E_k^2 + \omega i\eta}.\end{aligned} \tag{65}$$

The principle value of the integral has been previously calculated.[2,12,19]

Now, consider the following equation.

$$(-\frac{1}{c^2}\frac{\partial^2}{\partial t^2} - H)\psi^R(r,t) = f. \tag{66}$$

Its solution can be formally written as

$$\psi^R(r,t) = (-\frac{1}{c^2}\frac{\partial^2}{\partial t^2} - H)^{-1} f$$

$$= \int d\mathbf{r}' \int dt' (-\frac{1}{c^2}\frac{\partial^2}{\partial t^2} - H)^{-1} \frac{\partial \theta}{\partial t} \delta(\mathbf{r}-\mathbf{r}') f(\mathbf{r}',t') \qquad (67)$$

$$= \int d\mathbf{r}' \int dt' G^R(\mathbf{r},\mathbf{r}';t,t') f(\mathbf{r}',t').$$

In (67) the possible term $\varphi$ has been dropped which satisfies the homogeneous equation $(-\frac{1}{c^2}\frac{\partial^2}{\partial t^2} - H)\varphi = 0$. The equation (66) itself does not embody the time direction. The solution obtained through the procedure of (67) is endowed with the retard meaning. The factor $\theta(t-t')$ in the $G^R(\mathbf{r},\mathbf{r}';t,t')$ sets the actual upper limit of the time integration to be $t$ in (67).

In the case of an electromagnetic field, the retarded GF $G^R(\mathbf{r},\mathbf{r}';t,t')$ is (62). When (62) is substituted into (67), we have

$$\psi^R(\mathbf{r},t) = \int d\mathbf{r}' \int dt' \theta(t-t') \frac{f(\mathbf{r}',t')}{4\pi\varepsilon_0 |\mathbf{r}-\mathbf{r}'|} \delta(t-t'-\frac{|\mathbf{r}-\mathbf{r}'|}{c}). \qquad (68)$$

This is Eq. (6.65) in [10].

Suppose that starting from $t=0$, an electric charge $Q$ emerges at the origin $\mathbf{r}=0$,

$$f(\mathbf{r},t) = Q\theta(t)\delta(\mathbf{r})/\varepsilon_0. \qquad (69)$$

Then, substitution of (69) into (68) results in

$$\psi^R(\mathbf{r},t) = \theta(t) \frac{Q}{4\pi\varepsilon_0 |\mathbf{r}|} \theta(t-\frac{|\mathbf{r}|}{c}). \qquad (70)$$

Starting from the instant $t = 0$, a retarded potential begins to be established. The potential at the position $\mathbf{r}$ at time $t$ is emitted at the time $t-|\mathbf{r}|/c$ by the charge at the origin. Outside of the distance, the potential is zero. The time step function $\theta(t-t')$ in (68) is from (62) and plays a role in ruling out the time-advancing term, e.g., the second term in (61).

Let us briefly review a typical treatment, say, Section 6.5 in [10], where the equation is actually (3). Here, we denote the solution of Eq. (3) by $U(\mathbf{r},\mathbf{r}';t,t')$ and its Fourier component is denoted by $u(\mathbf{k};\omega)$. It was solved that

$$u(\mathbf{k};\omega) = \frac{1}{(2\pi)^4} \frac{1}{k^2 - \omega^2/c^2}. \qquad (71)$$

However, when $\omega^2 = c^2 k^2$, Eq. (71) was invalid and the inverse Fourier transformation of (71) could not be taken, and as a result, there is no solution for (3). To overcome the difficulty, the analytical continuation (6) was taken.

$$u(\boldsymbol{k};\omega) = \frac{1}{(2\pi)^4} \frac{1}{k^2 - \omega^2/c^2} \to u^+(\boldsymbol{k};\omega) = \frac{c^2}{(2\pi)^4} \frac{1}{c^2 k^2 - (\omega + i\eta)^2}. \tag{72}$$

It is easily calculated that the inverse Fourier transformation of $u^+(\boldsymbol{k};\omega)$ is

$$U^R(\boldsymbol{r},\boldsymbol{r}';t,t') = \frac{1}{(2\pi)^4} \int d\boldsymbol{k} \int d\omega e^{i\boldsymbol{k}\cdot(\boldsymbol{r}-\boldsymbol{r}')} e^{-i\omega(t-t')} u^+(\boldsymbol{k};\omega)$$
$$= \frac{ci\theta(t-t')}{(2\pi)^4} \int d\boldsymbol{k} e^{i\boldsymbol{k}\cdot(\boldsymbol{r}-\boldsymbol{r}')} \frac{\sin ck(t-t')}{k} = \theta(t-t')G(\boldsymbol{r},\boldsymbol{r}';t,t'). \tag{73}$$

In our opinion, $U^R(\boldsymbol{r},\boldsymbol{r}';t,t') = G^R(\boldsymbol{r},\boldsymbol{r}';t,t')$ is the retarded GF, but the $U(\boldsymbol{r},\boldsymbol{r}';t,t')$ is not. $U^R(\boldsymbol{r},\boldsymbol{r}';t,t')$ satisfies

$$(-\frac{1}{c^2}\frac{\partial^2}{\partial t^2} + \nabla_r^2)G^R(\boldsymbol{r},\boldsymbol{r}';t,t') = -\frac{\partial\theta(t-t')}{\partial t}\delta(\boldsymbol{r}-\boldsymbol{r}') \tag{74}$$

instead of (3). It is seen, therefore, that the analytical continuation (6) actually covertly replaces Eq. (3) with (74). A manipulation equivalent to the continuation (6) involved omitting the second term in (61),[11,16,19] which in fact was multiplied by a factor $\theta(t-t')$, as in Eq. (62).

The Fourier component of $G$ is denoted by $g$. Then, the Fourier component of $G^R$ is the convolution of the two Fourier components of the two factors in (55).

$$G^+(\boldsymbol{r},\boldsymbol{r}';\omega) = \frac{1}{2\pi}\int_{-\infty}^{\infty} d\omega' \frac{g(\boldsymbol{r},\boldsymbol{r}';\omega')}{\omega - \omega' + i\eta}. \tag{75}$$

The time Fourier transformation of (59) is

$$g(\boldsymbol{r},\boldsymbol{r}';\omega) = c\sum_n \varphi_n(\boldsymbol{r})\varphi_n^*(\boldsymbol{r}')\frac{\delta(\omega - E_n c) - \delta(\omega + E_n c)}{2i\sqrt{E_n}}. \tag{76}$$

Subsequently,

$$G^+(\boldsymbol{r},\boldsymbol{r}';\omega) = \sum_n \varphi_n(\boldsymbol{r})\varphi_n^*(\boldsymbol{r}')(\frac{1}{\omega - E_n c + i\eta} - \frac{1}{\omega + E_n c + i\eta}). \tag{77}$$

The equation that the advanced GF satisfies is

$$(-\frac{1}{c^2}\frac{\partial^2}{\partial t^2} - H)G^A(\boldsymbol{r},\boldsymbol{r}';t,t') = \frac{\partial\theta(t'-t)}{\partial t}\delta(\boldsymbol{r}-\boldsymbol{r}'). \tag{78}$$

The solution is of the form

$$G^A(\boldsymbol{r},\boldsymbol{r}';t,t') = -\theta(t'-t)G. \tag{79}$$

The initial conditions of $G$ are as follows.

$$G|_{t=t'} = 0. \tag{80a}$$

$$\frac{\partial G}{\partial t}\bigg|_{t=t'} = c^2 \delta(\mathbf{r} - \mathbf{r}'). \tag{80b}$$

In (80b), the time $t$ approaches $t'$ on the $t < t'$ side. When (79) is substituted into (78), we get

$$(-\frac{1}{c^2}\frac{\partial^2}{\partial t^2} - H)G = 0. \tag{81}$$

This $G$ is exactly the same as that in (55). For example, the auxiliary GF $G$ of the electromagnetic field is (61), and the corresponding advanced GF is

$$G^A = \theta(t'-t)\frac{1}{4\pi|\mathbf{r}-\mathbf{r}'|}\delta(t-t'+\frac{|\mathbf{r}-\mathbf{r}'|}{c}). \tag{82}$$

Similarly to the case of the first order equation, for the second order equation, there are two definitions of the Feynman propagator $\Delta_F$ in the literature. One definition is the solution of Eq. (3) or (4). The other is by means of the creation and annihilation operators, and it is the same as the causal GF in many-body theory. Some authors[8,15] have presented both definitions and considered their results to be identical.

From the discussion above, we know that when the solutions of (3) and (4) are Fourier transformed, as Eq. (5), the analytical continuation is taken to let the frequency be below the real $\omega$ axis, as in (6). This process is actually equivalent to turn (2) into (54). The continuation can also be taken to above the real axis, which is equivalent to convert the right hand side of Eq. (2) to be that of (78). Therefore, the poles of the $G^R(\omega)$ ($G^A(\omega)$) are inevitably below (above) the real axis.[6]

There is no such differential equation whose solution includes two terms each having a pole, with one pole below and the other above the real axis. Hence, in mathematical physics, there is no way to define a differential equation that the causal GF satisfies. The discussion is the same as the case of the first order equation in Section II. The analytical continuation (7) is incorrect.

## IV. DISCUSSION

We have been clear that in mathematical physics there are auxiliary GF $G$ and retarded and advanced GFs $G^R(\omega)$ and $G^A(\omega)$. Their relationships are Eqs. (25), (45), (55), and (79).

The integral path of frequency is always along the real axis, plus a possible semi-infinite circle in the lower or upper plane to compose a closed loop if necessary.

In the literature, closed integral paths other than the one along the real axis are found,[5,15,17,19] and some auxiliary functions, usually called invariant functions, are defined, each corresponding to a specific integral path.[19] These closed paths are not associated with rigorous mathematical proofs. Their accuracy is questionable because the poles were assumed to be on the real axis, but this is not true. As the $G$, $G^R$, and $G^A$ are given, there is probably no need for the other so-called invariant functions in mathematical physics.

In summary, there is no way to define the causal GF in mathematical physics. It is the analytical continuation of (6), but not (7), that can be rigorously verified as in the present work.

Here, we provide one physical reason why the causal GF cannot be defined in mathematical physics. It was mentioned above that among the many-body GFs, the simplest case is the average in the vacuum state of the creation and annihilation operator pairs of a noninteracting system, which was called one-body GF. Although there is no interaction between particles, the system is still a many-body system, say an ideal gas, so that it has a temperature. Among the one-body GFs, the retarded and advanced GF do not rely on temperature, whereas the causal GF does.[22] By contrast, in the differential equations listed in the present paper, the $H$ is a one-body operator. That is to say, we always treat the one-body problem such that there are no concepts such as ensemble, temperature, and so on. This is why there is no way to define the causal GF in mathematical physics.

The present paper is of two meanings. On one hand, we have clarified above the explicit differential equations that the retarded and advanced Green's functions satisfy in mathematical physics. On the other hand, this paper builds a foundation for our later work concerning QED and the problem of motion irreversibility.

We have pointed out above that some auxiliary functions defined by specific integral paths are incorrect. Equation (53) is a typical equation used in QED. We intend to enter the field of QED later and will then reexamine the problems involving the famous Feynman propagator.

Please note an important discrepancy between Eqs. (1) and (24). The former is of invariances of time translation and time inversion, but the latter is not. The reason is that the former is simply an equation of motion, whereas the latter includes not only an equation but also the initial condition, as shown by Eqs. (27a) and (27b). We stress that this is the first time that the initial condition is contained in the differential equation. The significance doing so is as follows. It is well known that the equations describing the motion of matters in the universe are differential equations that are of invariances of time translation and time inversion. People also know that to solve a concrete equation of motion, the initial conditions are unavoidable. The big bang theory demonstrates that the universe has a starting point of time. Consequently, no substance in the universe can move without initial conditions.

As long as the initial condition (27b) is included in the differential equation (24), the equation is not of time translation invariance. In other words, each concrete motion is not guaranteed to automatically have time translation invariance. This is because the motion has a starting point in time and does not cover the whole range of time. People usually think that each motion is of time translation invariance. This is because they pay close attention only to the differential equation regardless of the initial conditions.

For the same reason, people usually think that each motion is reversible in time, as the differential equation reveals, regardless of the initial conditions. Compared with Eq. (1), Eq. (24) is not of the invariance of time inversion, (please see Eq. (A4) in Appendix). The inclusion of the initial conditions in the differential equation opens the door to solving the problem of why there are irreversible processes. Equations (1) and (24) are for Green's functions. The corresponding equations of motion are expressed by (8) and (11), with the latter comprising the initial condition. In our future work, we will analyze cases where the motion is irreversible in time.

## V. CONCLUSIONS

In this paper, the differential equations that the retarded and advanced GFs satisfy are presented. The equations can be either the first or second order derivatives with respect to time. The skill in establishing the equations is that in the previous differential equations satisfied by the time-dependent GF, the time Dirac $\delta$ function is replaced by the derivative of the time step function. The key is that the result of taking the derivative of the time step is the Dirac $\delta$ function plus an infinitesimal term. The latter should not be omitted since it reflects the direction of the evolution of time. In this way, the retarded and advanced GFs are defined in terms of mathematical physics, since they can be directly solved from the corresponding differential equations.

In the frequency $\omega$ space, the poles of the retarded (advanced) GFs are always below (above) the real axis. The previously used analytical continuations of time-dependent GFs are no longer needed. It is stressed that any result ought to be rigorously mathematically derived. The artificial analytical continuation should be avoided as much as possible.

In mathematical physics, there is no way to define the causal GF because there is no corresponding differential equation.

**Acknowledgements**
This work is supported by the National Key Research and Development Program of China [Grant No. 2018YFB0704304].

**Appendix: The derivative of the step function**

The step function $\theta(x-x')$ is defined by[4,13,19]

$$\theta(x-x') = \begin{cases} 1, & x > x' \\ 1/2, & x = x' \\ 0, & x < x' \end{cases} \quad (A1)$$

The Dirac $\delta$ function is defined as follows.[24,32–37]

$$\delta(x) = 0, x \neq 0. \quad (A2a)$$

$$\int_{-\infty}^{\infty} \delta(x)dx = 1. \quad (A2b)$$

$$\int_{-\infty}^{\infty} f(x)\delta(x)dx = f(0). \quad (A2c)$$

Some properties of the $\delta(x-x')$ can be seen in [24,32–37]. A commonly used expression in the form of Fourier transformation is[24,32,34–37]

$$\delta(x-x_0) = \frac{1}{2\pi} \int_{-\infty}^{\infty} e^{-ik(x-x_0)} dk. \quad (A2d)$$

It is normally believed that the derivative of the step function is simply the Dirac $\delta$ function,[1,4,5,8,32]

$$\frac{\partial}{\partial x}\theta(x-x') = \delta(x-x'). \quad (A3)$$

In two textbooks,[22,24] it was particularly pointed out that

$$\frac{\partial}{\partial x}\theta(x-x') = \delta(x-x') - 0^+ \theta(x-x'). \quad (A4)$$

The second term in (A4) is an infinitesimal. It can be abandoned if no further derivation is carried out. It has to be retained, however, if the derivation continues. This infinitesimal comes from the Fourier transformation of the step function,[4,5,7,8,22]

$$\theta(x-x') = \frac{i}{2\pi} \int_{-\infty}^{\infty} \frac{e^{-ik(x-x')}}{k + i0^+} dk. \quad (A5)$$

In (A5), the infinitesimal in the denominator remains, because the integration is to be implemented. The infinitesimal determines that the pole position of the integrand is below the real axis, and is vital for the integration. The derivative of (A5) is

$$\frac{\partial}{\partial x}\theta(x-x') = \frac{i}{2\pi} \int_{-\infty}^{\infty} \frac{-i(k+i\eta-i\eta)e^{-ik(x-x')}}{k+i\eta} dk$$
$$= \frac{1}{2\pi} \int_{-\infty}^{\infty} e^{-ik(x-x')} dk - \eta \frac{i}{2\pi} \int_{-\infty}^{\infty} \frac{e^{-ik(x-x')}}{k+i\eta} dk, \eta \to 0^+ \quad (A6)$$

which leads to (A4).

We give an explanation, by comparing the two terms on the right-hand side of (A4), for why the infinitesimal in this equation must be kept. As $x = x'$, the first term is nonzero and, so, the second term can be discarded; as $x \neq x'$, the first term is zero, and although the second term is an infinitesimal, it must be retained when other terms are zero.

This reason is in fact the same as that of the following formula.

$$\frac{1}{x \pm i0^+} = P\frac{1}{x} \mp i\pi\delta(x), \tag{A7}$$

where $P$ means taking principle value. In the denominator of (A7), no matter how small the $x$ is, the imaginary part can be neglected; the imaginary part remains only when $x = 0$. People notice the necessity of keeping the infinitesimal in (A7) because of two reasons. One is that it is apparently distinguishable compared to the real part, and the other is that it is in the denominator and may therefore lead to infinity, as shown on the right-hand side of (A7). By contrast, both terms on the right hand side of (A4) are real and in the numerator, which makes the infinitesimal term not eye-catching.

The infinitesimal in (A4) comes from that in the integrand in (A5), i.e., (A7). Since the infinitesimal in (A7) must be retained, so too should that in (A4).

We turn to the integration of (A4), with the integral constant neglected. On the one hand,

$$\int_{-\infty}^{x} dx'' \frac{\partial}{\partial x''}\theta(x''-x') = \theta(x-x'). \tag{A8}$$

On the other hand,

$$\int_{-\infty}^{x} dx''[\delta(x''-x') - 0^+\theta(x''-x')] = \int_{-\infty}^{x} dx\delta(x''-x') = \theta(x-x'). \tag{A9}$$

In (A9), the term of the infinitesimal is neglected. This is because when $x \geq x'$, $\theta(x-x')$ is a nonzero figure, and the infinitesimal can be discarded. Hence, for the integral of the $\delta(x-x')$ function, the infinitesimal can be totally dropped.

Another step function is[7,9,24]

$$\theta(x'-x) = \frac{1}{2\pi i}\int_{-\infty}^{\infty} \frac{e^{-ik(x-x')}}{k-i0^+} dk. \tag{A10}$$

Its derivative is

$$\frac{\partial}{\partial x}\theta(x'-x) = -\delta(x-x') + 0^+\theta(x'-x). \tag{A11}$$

In the discussion in [5], the infinitesimal in (A11) was dropped. We stress that the infinitesimals in (A4) and (A11) have of physical meanings in the cases of time step function. They determine that the motion is respectively toward the future and past.

Finally, let us try to understand why the infinitesimal appears in Eq. (A4). Dirac $\delta$ function is investigated by means of the sophisticated theory of distribution, or generalized functions.[24,32–37] The property (A2c) can be rigorously developed as the limit of a sequence of functions, a distribution, and some examples are given.[24,32-37]

Now we inspect Eq. (A4). The term of infinitesimal is not from the Dirac $\delta$ function itself, but from the derivative of the step function. Therefore, we must examine the step function. The step function, usually not carefully investigated in the theory of distribution, may also be expressed as the limit of a sequence of functions. There are sequences the limits of which are just Eq. (A1). The followings are three examples, and their derivatives are simply the sequences for the Dirac $\delta$ function.

$$\theta_\eta(x) = \frac{1}{\pi}\tan^{-1}\frac{x}{\eta} + \frac{1}{2}, \quad \eta \to 0^+. \tag{A12}$$

$$\frac{d\theta_\eta(x)}{dx} = \delta_\eta(x) = \frac{1}{\pi}\frac{\eta}{\eta^2 + x^2}, \quad \eta \to 0^+. \tag{A13}$$

$$\theta_\eta(x) = \begin{cases} \frac{1}{2}e^{x/\eta}, & x \leq 0 \\ 1 - \frac{1}{2}e^{-x/\eta}, & x > 0 \end{cases} \quad \eta \to 0^+. \tag{A14}$$

$$\frac{d\theta_\eta(x)}{dx} = \delta_\eta(x) = \frac{e^{-|x|/\eta}}{2\eta}, \quad \eta \to 0^+. \tag{A15}$$

$$\theta_\eta(x) = \begin{cases} 0, & x < -\eta/2 \\ x/\eta + 1/2, & -\eta/2 < x < \eta/2 \\ 1, & x > \eta/2 \end{cases} \quad \eta \to 0^+. \tag{A16}$$

$$\frac{d\theta_\eta(x)}{dx} = \delta_\eta(x) = \begin{cases} 0, & |x| > \frac{\eta}{2} \\ \frac{1}{\eta}, & |x| \leq \frac{\eta}{2} \end{cases} \quad \eta \to 0^+. \tag{A17}$$

In these formulas, we intentionally let the parameters approach their limits $\eta \to 0^+$ in a uniform way.

The common features of the $\theta_\eta(x)$ are that

$$\lim_{\eta \to 0^+} \theta_\eta(x > 0) = 1 \tag{A18a}$$

and

$$\lim_{\eta \to 0^+} \theta_\eta(x < 0) = 0. \tag{A18b}$$

It is difficult to extract the information of an infinitesimal from any function in one sequence of $\theta_\eta(x)$ because it is finite everywhere. We explore the information from the Fourier transformation of $\theta_\eta(x)$.

The Fourier transformation (A5) is the exact expression of the step function, which results in the infinitesimal in (A4). It is expected that if any of the functions $\theta_\eta(x)$ in (A12), (A14), and (A16) is subject to Fourier transformation, the Fourier component at the limit $\eta \to 0^+$ should be $\dfrac{i}{k+i0^+}$ as revealed by (A5). Let us conduct the Fourier transformation.

The inverse Fourier transformation of (A2d) is

$$\int_{-\infty}^{\infty} e^{ik(x-x_0)}\delta(x-x_0)dx = 1. \tag{A19}$$

Now we have a pair of Fourier and inverse Fourier transformations for the Dirac $\delta$ function, (A2d) and (A18), and the Fourier transformation for the step function (A5). What remains to be derived is the inverse Fourier transformation for the step function.

Personally, since both the step function and Dirac $\delta$ function can be written as the limits $\eta \to 0^+$ of corresponding sequences of functions, we consider that the Fourier transformation should also be taken under the limit. Therefore, the Fourier transformation of a distribution $F_\eta(x)$ should be

$$\lim_{\eta \to 0^+} F_\eta(k) = \lim_{\eta \to 0^+} \int_{-\infty}^{\infty} e^{(ik-\eta)x} F_\eta(x) dx. \tag{A20}$$

Thus, the Fourier transformation of the sequence for the Dirac $\delta$ function is

$$\lim_{\eta \to 0^+} \int_{-\infty}^{\infty} e^{(ik-\eta)(x-x_0)} \delta_\eta(x-x_0) dk = 1. \tag{A21}$$

Equations (A13), (A15) and (A17) observe (21). Now, let us see the Fourier transformation of $\theta_\eta(x)$. By use of integration by parts and (A21),

$$\begin{aligned}
&\lim_{\eta \to 0^+} \int_{-\infty}^{\infty} e^{(ik-\eta)x} \theta_\eta(x) dx \\
&= \lim_{\eta \to 0^+} [\frac{e^{(ik-\eta)x}\theta_\eta(x)}{ik-\eta}]_{-\infty}^{\infty} - \frac{1}{ik-\eta} \lim_{\eta \to 0^+} \int_{-\infty}^{\infty} e^{(ik-\eta)x} \delta_\eta(x) dx \\
&= \frac{i}{k+i0^+},
\end{aligned} \tag{A22}$$

The first term in the second line of (A22) is zero because both the upper and lower limits are zero.

$$\lim_{\eta \to 0^+}[\frac{e^{(ik-\eta)x}\theta_\eta(x)}{ik-\eta}]_{-\infty}^{\infty} = \lim_{\eta \to 0^+}\lim_{x \to \infty}\frac{e^{(ik-\eta)x}\theta_\eta(x)}{ik-\eta} - \lim_{x \to -\infty}\lim_{\eta \to 0^+}\frac{e^{(ik-\eta)x}\theta_\eta(x)}{ik-\eta} = 0. \tag{A23}$$

That the upper limit is zero is guaranteed by (A18a) and the factor $e^{-\eta x}$, and that the lower limit is zero is guaranteed by (A18b). Thus, we complete the proof that the Fourier component of the step function is indeed $\dfrac{i}{k+i0^+}$.